# A Probabilistic Algorithm for Reducing Broadcast Redundancy in Ad Hoc Wireless Networks

**Md. Nazrul Islam, M.M.A. Hashem and A.M. Moshiur Rahman**
Department of Computer Science and Engineering
Khulna University of Engineering & Technology ( KUET )
Khulna 920300, Bangladesh.
kuet_nazrul2k@yahoo.com, hashem@cse.kuet.ac.bd, al_sonet@yahoo.com

**Abstract**

*In a wired network, a packet can be transmitted to a specified destination only, no broadcasting required. But in ad hoc wireless network a packet transmitted by a node can reach all neighbors due to broadcasting. This broadcasting introduces unnecessary retransmissions of same message. Therefore, the total number of transmissions (forward nodes) is generally used as the cost criterion for broadcasting. The problem of finding the minimum number of forward nodes is NP-complete. In this paper, the goal is to reduce the number of forward nodes which will reduce redundant transmission as a result. Thus some of approximation approaches are analyzed, especially dominant pruning and total dominant pruning which use 2-hop neighborhood information and a new approach: Probability based algorithm is proposed with a view to minimizing number of forward nodes. Simulation results of applying this algorithm shows performance improvements with compared to dominant pruning and total dominant pruning algorithms.*

## I. INTRODUCTION

One wireless network architecture that has attracted a lot of attention recently is the ad hoc wireless network [1]. In areas where there is a little or no communication, infrastructure is inconvenient to use, through the formation of an ad hoc wireless network user's still be able to communicate. An ad hoc wireless network is a collection of wireless mobile hosts forming a temporary network without the aid of any centralized administration or standard support services regularly available on the wide area network to which the host may normally be connected. Broadcasting process has extensive application in ad hoc wireless networks. The way that packets are transmitted in ad hoc wireless networks is quite different than the way that those are transmitted in wired networks. The significant difference is that, when a host sends a packet, all of its neighbors will receive that packet i.e. each node operates under the promiscuous receive mode. Therefore, the total number of transmission (forwards nodes) is generally used as the cost criterion [1]. This cost criterion also refers to time requirements, resource utilization, redundant message route lifetime, routing overhead etc.

A straight forward approach for broadcasting is *blind flooding,* in which each node will be obligated to re-broadcast the packet whenever it receives the packet for the first time. It may generate redundant transmissions. Figure 1 shows a network with three nodes. When node $u$ broadcasts a packet, both nodes $v$ and $w$ will receive the packet. Then $v$ and $w$ will rebroadcast the packet to each other.

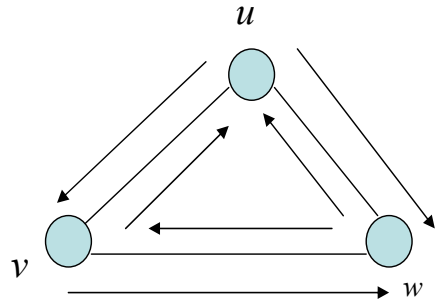

Fig. 1 Redundant transmission by blind flooding.

Apparently the last two transmissions are unnecessary. Such transmissions are called broadcast redundancy. Many broadcast algorithms besides blind flooding have been proposed. These algorithms utilize neighborhood and/or past information to reduce redundant transmissions. The *dominant pruning* algorithm and *total dominant pruning algorithm* are promising approaches those utilizes 2-hop neighborhood information to reduce redundant transmissions [1].

In this paper, a new broadcast technique has been proposed based on 1-hop neighborhood information. An algorithm based on this technique utilizes neighborhood information more effectively, termed as Probabilistic algorithm. Simulation results of applying this algorithm shows performance improvement with compared to existing *Dominant pruning* and *Total dominant Pruning* algorithms used for broadcasting in ad hoc wireless network. In addition, a termination criterion is discussed. The rest of the paper is organized as follows: Section II discusses some related works on reducing broadcast redundancy. Details about Dominant Pruning and Total dominant Pruning are also presented. An algorithm based on probability approach is given in Section III, a termination criteria is also discussed. Simulation results are shown in Section IV. Finally, section V concludes the paper and outlines future works.

## II. RELATED WORKS

Several Works & studies are performed for broadcasting in Ad Hoc wireless networks. These works are described as follows: - *Lim & Kim* provide two approximation algorithms: self pruning & dominant pruning. In [3], *Stojmenovic* studied a connected-dominant-set-based broadcast algorithm that uses only internal nodes to forward the broadcast packet. Internal nodes are dominating nodes derived by Wu and Li's *marking process* [3]. *Calinescu* [4] proposed a location-aware pruning method that extends the work of Lim and Kim. It is shown that the resultant dominating set has a constant approximation ratio of 6. In this paper, it is assumed that each host has no location information of other hosts and we will compare with only those protocols that do not depend on location information. Among the works performed for reducing broadcast redundancy in Ad hoc wireless networks, the efficient and attractive works are dominant pruning and total dominant pruning. Here, a simple graph, G =(V, E) is used to represent an ad hoc wireless network, where V represents a set of wireless mobile hosts (nodes) and E represents a set of edges. An edge $(u,v)$ indicates that both $u$ and $v$ are within their transmitter ranges. The circle around a host $u$ corresponds to the transmitter range of host $u$. All the hosts in the circle are considered the neighbors of host $u$. A host can obtain its neighborhood information by periodically sending an update message. Here, $N(u)$ represents the neighbor set of $u$ (including $u$). $N(N(u))$ represents the neighbor set of $N(u)$. The efficient algorithms uses criterion described above. Thus algorithms are noted down below:

### A. The dominant pruning (DP) Algorithm [1]

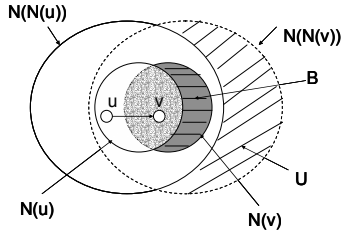

Fig. 2 Illustration for Dominant Pruning (DP).

### A.1 Dominant Pruning (DP) algorithm [1]

1) Node $v$ uses $N(N(v)), N(u)$ and $N(v)$ to obtain -
$U(u,v) = N(N(v)) - N(u) - N(v)$ and
$B(u,v) = N(v) - N(u)$.

2) Node $v$ then calls the selection process to determine $F(u,v)$. Here $F(u,v)$ represents the forward nodes' list.

### A.2 Selection process [1]

1) Let $F(u,v) = [\ ]$ (*empty*), $Z = \varphi$ (*empty*) and $K = \cup S_i$ where $S_i = N(v_i) \cap U(u,v)$ for $v \in B(u,v)$.

2) Find set $S_i$ whose size is maximum in $K$. (In case of a tie, the one with the smallest id $i$ is selected.)

3) $F(u,v) = F(u,v) \| v$, $Z = Z \cup S_i$, $K = K - S_i$. and $S_j = S_j - S_i$ for all $S_j \in K$

4) *If* $Z = U(u,v)$, *exit*; *otherwise go to step* 2.

As indicated in [1], the DP algorithm shows a better performance compared with other flooding algorithms such as blind flooding and self-pruning. In the DP algorithm, when node v receives a packet from node u, it selects a minimum number of forward nodes that can cover all the nodes in $N(N(v))$. Among nodes in $N(N(v))$, u is the source node and nodes in $N(u)$ have already received the packet and nodes in $N(v)$ will receive the packet after v rebroadcasts the packet. Note that, $N(u)$ can be directly derived from $N(N(V))$, once node v knows the sender's identification of u. Therefore, v just needs to determine its forward node list $F(u,v)$ from $B(u,v) = N(v) - N(u)$ to cover nodes in $U(u,v) = N(N(v)) - N(u) - N(v)$. $U(u,v)$ is the area with oblique lines in the above figure. Specifically, the greedy set cover algorithm is used for the selection of forward nodes.

### B. The Total Dominant Pruning [1]

If node v can receive a packet piggybacked with $N(N(u))$ from node u, the 2-hop neighbor set that needs to be covered by $v's$ forward node list $F$ is reduced to
$$U(u,v) = N(N(v)) - N(N(u)).$$

The total dominant pruning algorithm uses the above method to reduce the size of *U* and, hence, to reduce the size of *F*. The region $N(N(v))$, $N(N(u))$, $N(v)$ and $N(u)$ are shown in the figure below.

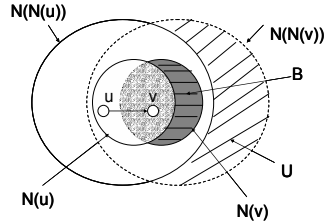

Fig. 3 Illustration for Total Dominant Pruning (TDP).

**B.1 Total dominant Pruning (TDP) Algorithm:**

1) Node $v$ uses $N(N(v)), N(u)$ and $N(v)$ to obtain
$U(u,v) = N(N(v)) - N(N(u))$ and
$B(u,v) = N(v) - N(u)$.

2) Node $v$ then calls the selection process described above to determine $F(u,v)$. Here $F(u,v)$ represents the forward nodes' list.

## III. PROPOSED PROBABILITY BASED ALGORITHM

In this section, a proposal for broadcast using 1-hop neighborhood information is introduced first and then explained with example.

### A. The approach

To cover the whole network using few numbers of nodes and avoiding redundant transmissions of message advances us to the following condition –

- A node is to be selected as forward if the node has maximum number of neighbors.
- Region of nodes to be covered by a node should be provided. It refers to the records of uncovered nodes in the network.

Thus a "Probability" based algorithm is proposed in which a node is selected as forward node on the basis of its probability value. This "Probability" is the probability of disheartening related source to cover an area. If a node has minimum probability value then it assures the related source to cover maximum region expected by the source. Based on probability value a node is selected by the related source if its value is minimum. Thus such "Probability" based algorithm depends upon several calculations and make decisions based on these calculations.

### A.1 Probability calculation

In probability based algorithm, initially the probabilities of all nodes are initialized to zero. Whenever a node is to transmit a message that is acting as source, it then calculates the probability of each of its neighbors. The process of calculating probability may be expressed in the following form:

$$P_i = 1/n_i \qquad (1)$$

Here, $P_i$ = Probability of node i and $n_i$= Number of nodes that may be covered by node i.

$n_i$ = (Neighbors of node i – Neighbors of relative source) $\cap$ U  (2)

Moreover, nodes to be covered are listed in *Uncovered List*. This list is denoted by U. Nodes already covered by a node are stored in *Covered List* which will be denoted by $C_i$. Probability calculation is explained by the example given below-

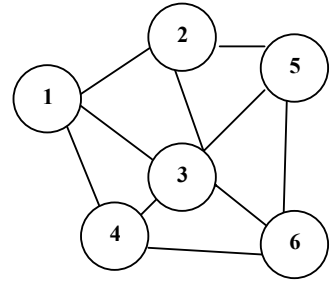

Fig. 4   A simple network model

Let us assume node 1 as the source. It initializes as follows –

$U = \{1,2,3,4,5,6\} - N(Source) = \{5,6\}$

Value of $n_i$ for each neighbor i of the source is calculated as follows –

Refer to (2) $n_2$ =1. Similarly, $n_3 = 2$ and $n_4 = 1$. Probabilities are calculated refer to (1) as $P_2 = Inverse\ of\ n_2 = 1/1 = 1, P_3 = 1/n_3 = 1/3$, $P_4 = 1/1 = 1$. Thus node 3 has smallest value of probability. Hence node 3 is selected as forward node. Source 1 sends the forwarding capability to node 3 with U{5,6}. Node 3 sets its probability as $P_3 = \infty$. Whenever node 3 covers node 5 & 6, it then sends information to its neighbors about the nodes covered by itself and probability of self i.e. $P_3 = \infty$. Thus probability of forward node 3 is globally updated. Then source 1 calculated probabilities as follows:   U={}, $P_3 = \infty$ $n_2 = (\{5\} - N(1)) \cap U = 0$, $n_4 = 0$. Source 1 sets the probabilities as - $P_2 = \infty, P_3 = \infty\ \&\ P_4 = \infty$. Thus source 1 ends its sending forwarding capability to its neighbors and terminates its broadcasting.

### A.3 Probability updating

Whenever a node sends retransmission capabilities to one of its neighbors, it updates the probability of the neighbor by assigning infinite value. It's termed as local updating. Whenever a forward node completes its transmission, it sends its probability value assigned infinity to its neighbors. Thus all the neighbors are aware of its probability value. It's termed as global updating.

### A.4 Uncovered list

After completing of transmission a forward node updates its uncovered list as follows –

$U = U - C_f$  (3)

where $C_f$ refers to the *CoveredList* of forward node **f**.

It is local updating. It then sends covered-list information along with its own probability value to its neighbors. The neighbors then update their uncovered list as follows –

$U = U - C_f$.

It is global Updating.

### A.5 Probabilistic Algorithm

Phase 1: Initialize source.

Phase 2: Set U = all the nodes in the network – Neighbor of Source  // Initialization phase

Set $P_i = 0.0$ for each node i // initializing Probability value

/*This is the phase in which forward nodes are selected */

Phase 3:

  **i)** Source 1:=Source.

  **ii)** Source 1 calculates the probability of all its neighbors.

  **iii)** Source 1 selects the node with smallest probability as forward node. If no node is available to be selected as forward then go to step vii.

  **iv)** Source 1 sends uncovered list to the selected forward node.

  **v)** Source 1 adds the forward node into its forward list.

  **vi)** Source 1 updates its uncovered list with *covered-list* available from forward node. Go to step **ii**.

  **vii)** Forward pointer is incremented to show the next forward node of source 1. Swap the value of source 1 along with the current forward node available in its list. Go to step **ii**.

### B. Example

Let node 6 be the source in a network as given in figure 5. Then calculations performed are as follows:

- Source 6 calculates the probability of the neighbors as follows:
  U = {1,2,3,4,5,6,7,8,9,10,11,12} – N(*Source*) = {1,3,4,8,10,11,12}

Refer to (1) and (2), $n_7 = 3$, $P_7 = 1/n_7 = 1/3 = 0.333$, $n_2 = 2$,

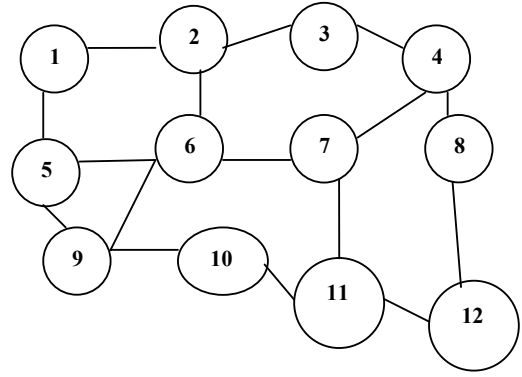

Fig. 5  A simple network of 12 nodes

$P_2 = 1/n_2 = 1/5 = 0.2$, $n_5 = 1, P_5 = 1/n_5 = 1/1 = 1$, $n_9 = 1 \& P_9 = 1/1 = 1$. So, source 6 selects node 7 as forward node as its probability is minimum. Source 6 sends rebroadcast capability with uncovered-list U to node 7. Node 7 covers nodes 4, 8 & 11. It sets its probability value to infinity. It sends its probability value and covered list to its neighbors.

- Source 6 updates its uncovered list as –
  U=U – C$_7$ = *{1, 3, 10, 12}*

It calculates probabilities again – $P_2 = 0.5, P_7 = \infty, P_6 = \infty$, $P_5 = 1 \& P_9 = 1$.

So, it selects node 2 as next forward. Node 2 covers nodes 1 & 3. It sets its probability value to infinity. It sends its probability value and covered list C$_2$ to its neighbors.

- Source 6 updates its uncovered list as –

  U= U – C$_2$ ={10,12}

It calculates probabilities again –

$P_2 = \infty, P_6 = \infty, P_7 = \infty$,     $P_5 = \infty$ **as**    $n_5 = 0$, $P_9 = 1$ **as** $n_9 = 1$ **&** $N(9) = \{10\}$

So, it selects node 9 as next forward. Node 9 covers node 10. It sets its probability value to infinity. It sends its probability value and covered list to its neighbors.

- Source 6 updates its uncovered list as –
  U = U – C$_9$

It calculates probabilities as – $P_2 = \infty, P_5 = \infty, P_6 = \infty, P_7 = \infty, P_9 = \infty$.

Source 6 then sends its source capability to node 7 (swap process) with *Uncovered list* **U**.

Source 7 updates its uncovered list as – **U = U – C$_6$**. It calculates probabilities as – $P_4 = \infty$ as $n_4 = 0$, $P_8 = 1$ as $n_8 = 1, N(8) = \{12\}$, $P_{11} = 1$ as $n_{11} = 1, N(11) = \{12\}$. In case of tie, source selects the node with minimum identification value of i. So, 7 selects node 8 as forward.

Node 8 covers node It sets its probability to infinity and sends its probability value & its covered list $C_8$ to its neighbors. Source 7 updates its uncovered list as – $U = U - C_8$. It calculates probabilities as follows – $P_4 = \infty, P_8 = \infty, P_{11} = \infty$. It stops its rebroadcast as all probability values are Infinity .Node 7 sends covered list received from its selected forward nodes (8) to its source 6. Source 6 updates its uncovered list as – $U = U - C_7 = \{\}$.

Thus it is assured that all the nodes are covered and terminates its broadcasting. Thus forward node list is = {6,7,2,9,8}.

**C. Termination Criteria**

When a source node broadcasts a packet, each intermediate node will decide whether to rebroadcast the packet or to drop it independently, based on a given termination criterion. In other words, the broadcast process at each node will terminate when a given termination criterion is satisfied. To determine a termination criterion that guarantees delivery, we assume the following "static" environment. Mobile hosts are still allowed to roam freely in the working space. However the broadcast process (including the forward node selection & the broadcast process itself) is done quickly so that N (*v*) & N (source 1) remains same during the process for each host *v*. In addition, each host *v* has updated and consistent $N(v)$ when the broadcast process starts. In proposed approach the termination criterion is based on following two conditions:

1) Whenever a node finds its *UncoveredList* U empty then it stops retransmission and terminates broadcast.

2) If a node finds its *UncoveredList* U non-empty and no more neighbor is found to select as forward node to cover remaining region, then it assumes that this remaining region is beyond of its coverage and terminates broadcast.

## IV. PERFORMANCE SIMULATIONS

We simulate the performance of the dominant Pruning, total dominant Pruning and Probabilistic algorithm in terms of the average number of forward nodes generated. The simulation is conducted assuming a static environment. The simulator designed randomly generates graph of different structure. The number of hosts ranges from 5 to 30. For each given number of hosts, random graphs are generated. To avoid contention or collision, an ideal MAC layer is assumed. The following figure shows the comparison of proposed method with dominant pruning and total dominant pruning algorithms for variable source. Graphs representing networks of different nodes are used for simulation. For each graph a random number of sources are used to determine number of forward nodes. Then the average value was taken and using these values the graph was plotted.

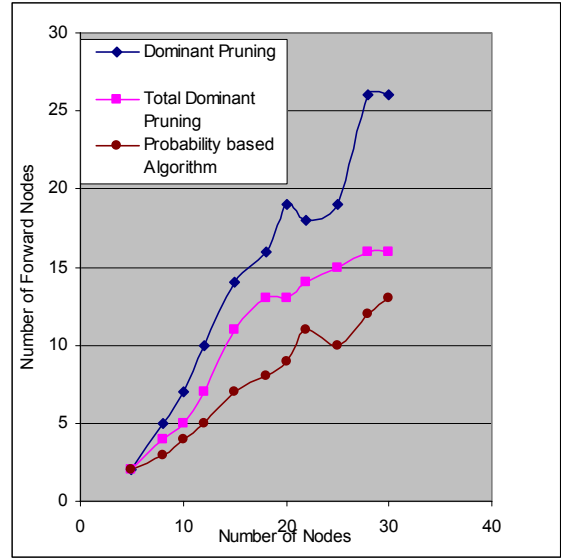

Fig. 6 Comparison of Proposed algorithm with "Dominant pruning" &"Total dominant pruning" for variable source.

From the figure above the following decisions can be made:

**1.** The proposed "Probability based algorithm" performs more efficiently than "Dominant pruning".

**2.** The proposed method performs efficiently than "Total Dominant Pruning".

The following figure shows the comparison of proposed method with dominant pruning and total dominant pruning algorithms using fixed source. The procedure stated for variable source was followed for this purpose. The only difference is that source was kept fixed for all different graphs representing network structure.

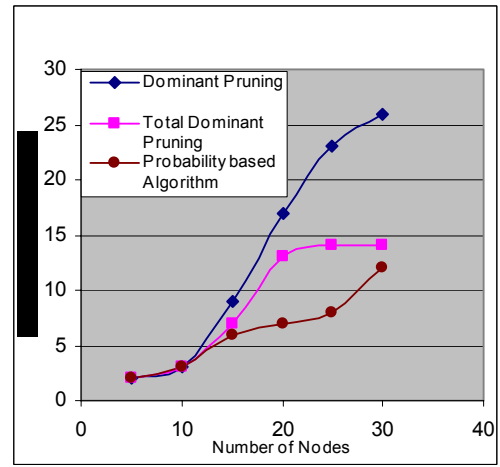

Fig. 7 Comparison of Proposed algorithm with "Dominant pruning" &"Total dominant pruning" for fixed source.

From the simulations (from Fig.6 & Fig. 7), we can see that differences among these algorithms exist in terms of broadcast redundancy. This is because the degree of broadcast redundancy directly relates to the number of the forward nodes. The more the number of forward nodes in a broadcast process, the higher the broadcast redundancy.

## V. CONCLUSIONS

In particular, the objective of this paper was to introduce a new technique for broadcasting in ad hoc wireless network & analyze the new approach with existing techniques. The technique which was named as "Probabilistic algorithm" is experimented with other approaches and according to the simulation results, it is observed that the proposed probability based algorithm obtains better solutions and works more efficiently. Moreover, it uses only 1-hop neighbor information for various calculations that results in less memory utilization and less processing delay. There are many ways in which the research work may be enhanced or expanded. The possible fields in which the studies can be performed in near future are as follows –

- The proposed technique can be applied and experimented more efficiently than the existing one with new features.
- The researchers may increase the neighborhood information from 1- Hop to 2- Hop to improve the performance of proposed approach.
- The probability value calculated using proposed approach and distance between nodes may be combined to determine heuristic value for each node which can be used to select forward node. The formula for calculating heuristic may be of the form-

   $h(n) = P(n) + D(n)$
   $h(n)$ = Heuristic value of node n
   $P(n)$ = Probability value of node n
   $D(n)$ = Distance or edge cost between nodes

The node having minimum heuristic will be selected as forward node.

## REFERENCES

[1]. Wei Lou & Jie Wu., "On Reducing Broadcast Redundancy in Ad Hoc Wireless Networks," *IEEE Transactions. on Mobile Computing*, vol 1. no.2 (2002), pp. 111-122.

[2] S. Ni, Y. Tseng, Y. Chen and J. Sheu, "The Broadcast Storm Problem in a Mobile Ad Hoc Network," *Procs. Of the MOBICOM '99,* pp. 151-162.

[3] W. Peng & X.C. Lu, "On the Reduction of Broadcast Redundancy in Mobile Ad Hoc Networks," *Procs .of First Ann Workshop Mobile and Ad Hoc Networking and Computing, MOBIHOC ,*2000, pp. 129-130.

[4] I. Stojmenovic, S. Seddigh & J. Zunic, "Dominating Sets and Neighbor Elimination Based Broadcasting Algorithm in Wireless Networks," *IEEE Trans. on Parallel and Distributed* System, vol.13, no.1, 1999, pp. 46-55.

[5] G. Calinescu, I. Mandoiu, P.J. Wan & A. Zellkovsky, "Selecting Forwarding Neighbors in Wireless Ad Hoc Networks", *Procs. Of the ACM International Workshop Discrete Algorithms and Methods for Mobile Computing* (DIALM '01), 2001, pp, 34-33.

[6] H. Lim & C. Kim "Flooding in Ad Hoc Wireless networks", *Computer Communications,*vol.24, no.3-4, 2001, pp. 353-363 .

[7] J. Wu & H. Li , "On Calculating Connected Dominating Sets for Efficient Routing in Ad Hoc Wireless Networks", *Procs. Of the ACM International Workshop Discrete Algorithms and Methods for Mobile Computing* , 2002, pp. 14-25.